\documentclass[journal,comsoc]{IEEEtran}

\usepackage[T1]{fontenc}
\usepackage{amsmath}
\usepackage{graphicx}
\usepackage{mathtools}
\usepackage{amsfonts}
\usepackage{amssymb}
\usepackage{relsize}
\usepackage{bm}
\usepackage{cite}
\usepackage{amsthm}
\usepackage{xcolor}
\usepackage{multirow}
\usepackage{hhline}
\DeclareMathOperator*{\argmin}{arg\,min}
\DeclareMathOperator*{\argmax}{arg\,max}
\ifCLASSINFOpdf
 \DeclareMathAlphabet\mathcal{OMS}{cmsy}{m}{n}
\DeclareMathAlphabet\mathbfcal{OMS}{cmsy}{b}{n}

\else

\fi

\newtheorem{The}{Theorem}
\newtheorem{Def}{Definition}
\newtheorem{Lem}{Lemma}
\newtheorem{Pro}{Proposition}

\begin{document}

\title{On the Design of Multi-Dimensional Irregular Repeat-Accumulate Lattice Codes}

\author{Min~Qiu,~\IEEEmembership{Student Member,~IEEE,}
        Lei~Yang,~\IEEEmembership{Member,~IEEE,}
        Yixuan~Xie,~\IEEEmembership{Member,~IEEE,}
        and~Jinhong~Yuan~\IEEEmembership{Fellow,~IEEE,}

\thanks{M. Qiu, L. Yang, Y. Xie and J. Yuan are with the School
of Electrical Engineering and Telecommunications, University of New South Wales, Sydney,
NSW, 2052 Australia (e-mail: min.qiu@student.unsw.edu.au; lei.yang3@unsw.edu.com; yixuan.xie@unsw.edu.au; j.yuan@unsw.edu.au). This paper was presented in part at the IEEE International Symposium on Information Theory (ISIT) in 2017 \cite{Qiu17}.}
}


\maketitle

\begin{abstract}
Most multi-dimensional (more than two dimensions) lattice partitions only form additive quotient groups and lack multiplication operations. This prevents us from constructing lattice codes based on multi-dimensional lattice partitions directly from non-binary linear codes over finite fields. In this paper, we design lattice codes from Construction A lattices where the underlying linear codes are non-binary irregular repeat-accumulate (IRA) codes. Most importantly, our codes are based on multi-dimensional lattice partitions with \emph{finite constellations}. We propose a novel encoding structure that \emph{adds} randomly generated lattice sequences to the encoder's messages, instead of multiplying lattice sequences to the encoder's messages. We prove that our approach can ensure that the decoder's messages exhibit permutation-invariance and symmetry properties. With these two properties, the densities of the messages in the iterative decoder can be modeled by Gaussian distributions described by a single parameter. With Gaussian approximation, extrinsic information transfer (EXIT) charts for our multi-dimensional IRA lattice codes are developed and used for analyzing the convergence behavior and optimizing the decoding thresholds. Simulation results show that our codes can approach the unrestricted Shannon limit within 0.46 dB and outperform the previously designed lattice codes with two-dimensional lattice partitions and existing lattice coding schemes for large codeword length.
\end{abstract}


%
\IEEEpeerreviewmaketitle

\section{Introduction}

\IEEEPARstart{L}{attices} are effective arrangements of equally spaced points in Euclidean space. They have attracted considerable attentions in the coding community because their appealing algebraic structures can be efficiently exploited for encoding and decoding. Although Shannon has shown that the optimal coding strategy to achieve Gaussian channel capacity is random coding with Gaussian distribution \cite{Shannon1948}, these random codes are more or less prohibited in practice. Lattice codes can be deemed as a natural alternative to random Gaussian codes. The remarkable work \cite{Zamir04} has proved the existence of lattice codes achieving the capacity of additive white Gaussian noise (AWGN) channels by using a lattice decoder. This decoder is suboptimal compared with the optimal maximum-likelihood (ML) decoder but has a lower decoding complexity. Apart from point-to-point communications, lattice codes have also been proved to be useful in a wide range of applications such as index coding \cite{Natarajan15}, cooperative communications \cite{Huang13}, multiple access \cite{Fang16}, multiple antenna systems \cite{Hindy17} and so on. It is believed that lattice codes will play a crucial role in future communication systems.

\subsection{Literatures and Motivations}
There is tremendous work on lattice codes which mainly include information theoretical analysis of their capacity achieving properties in different communication systems and lattice codes construction for practical systems. We focus on the code construction.

According to the literature, there are two main approaches to construct lattice codes. The first one is to construct lattice codes directly in the Euclidean space. There are two well-known examples: low-density lattice codes (LDLC) \cite{Sommer08} and convolutional lattice codes (CLC) \cite{Shalvi11}. Another approach is to adapt modern capacity approaching error correction codes to construct lattices, i.e., low-density parity-check (LDPC) lattices \cite{Sadeghi06,Pietro12,Tunali15} and polar lattices \cite{Yan13}. Their construction methods involve some well-known methods such as Construction A \cite{Conway99} (constructing lattices based on a linear code), Construction D \cite{Conway99} (constructing lattices based on the generator matrices of a series of nested linear codes), and Construction D' \cite{Conway99} (constructing lattices based on the parity check matrices of a series of nested linear codes). These methods allow one to construct lattice codes not only with good error performance inherited from capacity-achieving linear codes, but also having relatively lower construction complexity compared with LDLCs and CLCs. To sum up, most of the aforementioned designs have been shown to approach the Poltyrev limit \cite{Poltyrev94} (i.e., the channel capacity without either power limit or restrictions on signal constellations) within 1 dB when the codeword length is long enough. In addition, all of these lattices can be decoded with efficient decoding algorithms.

However, for LDLCs, in order to attain the best possible decoding performance, the decoder would have to take the whole probability density functions (pdf) for processing. This would require a significant amount of memory. As reported in \cite{Shalvi11}, the symbol error rate (SER) of the CLCs is higher than that of LDLCs. Both of these two lattice coding schemes are still difficult to implement in practice due to the use of non-integer lattice constellations. The LDPC lattices in \cite{Sadeghi06} and the polar lattices \cite{Yan13} involve multilevel coding and multistage decoding due to their construction methods. This poses a much higher complexity in encoding and decoding than that of Construct A lattices in \cite{Pietro12} and \cite{Tunali15}.

Since most of the available designs are based on infinite lattice constellations, their error performances are compared against Poltyrev limit. To put these lattice codes into practice, a power constraint must be satisfied. Moreover, most lattice codes have high complexity encoding structures due to the sparseness of their parity-check matrices which in general can lead to high-density generator matrices. Furthermore, most of the Construction A, Construction D and Construction D$^\prime$ lattice codes are designed based on one or two-dimensional (real dimension) lattice partitions. It is understood that this can result in a shaping loss in error performance compared with using higher-dimensional lattice partitions \cite{Shum15}. Constructing codes over multi-dimensional lattices have been investigated in \cite{Kositwattanarerk15,Oggier13,Kositwattanarerk13,Khodaiemehr16}. In \cite{Kositwattanarerk15} and \cite{Kositwattanarerk13}, the authors proposed a method for constructing lattices over number fields and have studied their application in wiretap block fading channels. In \cite{Oggier13}, the authors have proposed a lattice construction method to allow Construction A lattices equipped with multiplication, which has potential application in nonlinear distributed computing over a wireless network. In \cite{Khodaiemehr16}, the authors have designed lattices to obtain diversity orders in block fading channels. However, \cite{Kositwattanarerk15,Oggier13,Kositwattanarerk13,Khodaiemehr16} mainly focused on constructing lattices over algebraic number fields with applications to block fading channels while designing lattice codes to approach the unrestricted Shannon limit (i.e., when transmission is power limited but not restricted to any signal constellation) was not taken into account.

Recently, we have designed irregular repeat-accumulate (IRA) lattice network codes with finite constellations for two-way relay channels (TWRC) in \cite{Qiu16}. The lattice codes are constructed via Construction A on non-binary IRA codes. We have used the extrinsic information transfer (EXIT) charts to optimize the degree distribution in a bid to minimize the required decoding signal-to-noise ratio (SNR). However, this scheme is based on two-dimensional lattice partitions and thus still has a performance gap to the unrestricted Shannon limit.

\subsection{Problem Statement}
In light of the previous work, we aim to design multi-dimensional lattice codes to further approach the channel capacity. That being said, directly extending the design in \cite{Qiu16} to multi-dimensional lattice partitions is very challenging. There are two fundamental reasons why this is the case. First, in the previous setting, we employed a two-dimensional lattice partition to form a quotient ring which is isomorphic to a finite field. However, most multi-dimensional lattice partitions form additive quotient groups where addition is the only group operation. If we use multi-dimensional lattice partitions in our previous design, the multiplication between two lattice points cannot be performed on additive groups. Second, simply removing the multiplication in the encoding structure will prevent us from analysing and optimizing the multi-dimensional IRA lattice codes effectively. In the previous design, the encoder's messages are multiplied by some randomly generated sequences so that the permutation-invariant property \cite{Bennatan06} can be obtained. Under this property, the analysis and optimization of our lattice codes can be significantly simplified. It is possible to remove all the operations of multiplying random sequences to allow the use of multi-dimensional lattice partitions. However, the permutation-invariance property will not hold in this case. As a result, the densities of the messages in the iterative decoder can only be represented by a multivariate Gaussian distribution. This will lead to an extremely high complexity for our design and analysis.

\subsection{Main Contributions}
In this paper, we aim to design multi-dimensional IRA lattice codes with \emph{finite constellations} to further approach the unrestricted Shannon limit. This is different from most lattice codes which are based on infinite constellations in the literature. Even though these codes have been shown to approach the Poltyrev limit within 1 dB, it is still unclear whether these codes with power constraint can approach the unrestricted Shannon limit within 1 dB. In order to practically approach the unrestricted Shannon limit, we must optimize the degree distribution of our codes based on constellations, detection methods and decoding algorithms. Furthermore, we continue to use Construction A as it has been proved to be a simple and powerful tool for constructing capacity-achieving lattice codes \cite{diPietro16}. The main contributions of our work are summarized as below:
\begin{itemize}
\item We designed practical lattice codes with finite constellations based on multi-dimensional lattice partitions. More specifically, we proposed a novel encoding structure that adds random lattice sequences to the encoder's messages (output of the interleaver, combiner and accumulator). In addition, we introduced a constraint on the random lattice sequences in our encoder and proved that the constraint can lead to linearity of our codes. Since no multiplication is required in our encoder, our design can be directly applied to any lattices of any dimensions.

\item We investigated the optimal degree distributions of our lattice codes, aiming at approaching the unrestricted Shannon limit. We proved and showed that our encoding structure can produce permutation-invariant and symmetric effects in the densities of the decoder's messages (soft information propagated in the iterative docoder). These two properties enable to use a Gaussian distribution characterised by a single parameter to model the soft information propagated inside the iterative decoder. Under this condition, we used a two-dimensional EXIT chart to analyse the convergence behaviour of the iterative decoder. With EXIT charts, we designed a set of lattice codes for different target code rates with the minimum decoding threshold.

\item Numerical results are provided and show that our designed and optimised lattice codes can approach the unrestricted Shannon limit within 0.46 dB. We demonstrate that our lattice codes not only outperforms previously designed lattice codes in \cite{Qiu16} with two-dimensional lattice partitions, but also have less coding loss compared with the existing lattice coding schemes in \cite{Boutros16,Boutros14,diPietro16,Sommer06,Khodaiemehr17} for large codeword length, i.e., a codeword has more than 10,000 symbols.

\end{itemize}

\subsection{Structure of the Paper}
The rest of the paper is organised as follows. Section II provides some background knowledge of lattices and lattice codes. In Section III we present our lattice coding design including the construction of our lattice codes, the design of the encoder and decoder, as well as a design example of employing the $D_4$ lattice partition. Next in Section IV we explain how to model the soft information in the decoder. Most notably, we prove and show that our proposed lattice codes can achieve permutation-invariance and symmetry properties in the densities of the decoder's messages. The complete proof of all the theorems and lemmas is in Appendix. We also provide the convergence analysis by using EXIT chart and explain the use of EXIT chart curve fitting techniques to design our codes with optimal degree distributions in this section. The simulation results in Section V show the goodness of our proposed codes compared with the codes with two-dimensional lattice partitions. Finally, this paper finishes with Section VI summarizing our main achievements from this work.

\section{Background on Lattices and Lattice Codes}
In this section, we provide some essential definitions in relation to lattices \cite{Conway99} and lattice codes \cite{Zamir15}. All of these will be used throughout the rest of the paper. Note that all the concepts below are introduced based on real-dimensional lattices. Complex lattices can be defined in a similar manner as real lattices and thus will not be explicitly introduced here.

An $n$-dimensional \emph{lattice} $\Lambda$ is a discrete set of points $\boldsymbol{\lambda}$ in $\mathbb{R}^n$. It can be generated from an $n \times n$ full rank \emph{generator matrix} $\mathbf{G}_{\Lambda}$ with real entries which can be either integers or non-integers:
 \begin{equation}\label{eq:4b}
\Lambda = \{\boldsymbol{\lambda} = \mathbf{b} \mathbf{G}_{\Lambda}, \mathbf{b} \in \mathbb{Z}^n \},
\end{equation}
Note that we have restricted our definition to full-rank lattices because we do not need to treat lower-rank lattices for the purposes of this work. Here, $\boldsymbol{\lambda}$ is a lattice point with dimension $n$ or it can be deemed as a \emph{lattice codeword} with length $n$. All the lattices must contain the origin $\mathbf{0}$.

Lattices are \emph{groups} that are closed under addition:
\begin{equation}
\forall \boldsymbol{\lambda}_1, \boldsymbol{\lambda}_2 \in \Lambda, \; \boldsymbol{\lambda}_1 + \boldsymbol{\lambda}_2 \in \Lambda.
\end{equation}

The \emph{Voronoi region} associated with the lattice point $\boldsymbol{\lambda}$ is defined as:
\begin{equation}\label{eq:1}
\mathcal{V}_{\Lambda}(\boldsymbol{\lambda}) = \{\mathbf{x} \in \mathbb{R}^n, \; \; \|\mathbf{x} - \boldsymbol{\lambda} \| \leq  \|\mathbf{x} - \boldsymbol{\lambda^{\prime}} \|, \; \;  \forall \boldsymbol{\lambda^{\prime}} \in \Lambda \}.
\end{equation}
The \emph{fundamental Voronoi region} $\mathcal{V}_{\Lambda}(\mathbf{0})$ is the Voronoi region associated with the all-zero lattice point.

A \emph{lattice quantizer} or a \emph{lattice decoder} with respect to the lattice $\Lambda$ is denoted by $Q_{\Lambda}(\mathbf{x})$. It maps a point $\mathbf{x}$ in $\mathbb{R}^n$ to its closest lattice point:
\begin{equation}\label{eq:1a}
Q_{\Lambda}(\mathbf{x}) = \argmin_{\boldsymbol{\lambda} \in \Lambda}\| \mathbf{x} -  \boldsymbol{\lambda} \| .
\end{equation}

Recall the definition of the Voronoi region from above, if $\mathbf{x} \in \mathcal{V}_{\Lambda}(\boldsymbol{\lambda})$, then we have the following:
\begin{equation}\label{eq:1b}
Q_{\Lambda}(\mathbf{x}) =  \boldsymbol{\lambda} \in \Lambda.
\end{equation}

The \emph{modulo-lattice} operation is represented as:
\begin{equation}\label{eq:1c}
\mathbf{x} \;\text{mod}\; \Lambda = \mathbf{x} - Q_{\Lambda}(\mathbf{x}).
\end{equation}
It is the difference between a vector and its closest lattice point. So the output of this operation is always a point in the Voronoi region $\mathcal{V}_{\Lambda}(\boldsymbol{\lambda})$.

We denote the modulo-lattice addition with respect to $\Lambda$ by ``$\oplus$'' where
\begin{equation}\label{eq:1d}
\boldsymbol{\lambda}_1 \oplus \boldsymbol{\lambda}_2 = (\boldsymbol{\lambda}_1+\boldsymbol{\lambda}_2)\;\text{mod}\; \Lambda^\prime, \; \; \boldsymbol{\lambda}_1,\boldsymbol{\lambda}_2 \in \Lambda.
\end{equation}
Similarly, we define the modulo-lattice subtraction ``$\ominus$'' as follows:
\begin{equation}\label{eq:1dss}
\boldsymbol{\lambda}_1 \ominus \boldsymbol{\lambda}_2 = (\boldsymbol{\lambda}_1-\boldsymbol{\lambda}_2)\;\text{mod}\; \Lambda^\prime, \; \; \boldsymbol{\lambda}_1,\boldsymbol{\lambda}_2 \in \Lambda.
\end{equation}

A \emph{sublattice} $\Lambda^{\prime}$ of a lattice $\Lambda$ is a subset of the lattice $\Lambda$ that is a lattice itself. We say $\Lambda^{\prime}$ is \emph{nested} in $\Lambda$ if $\Lambda^{\prime} \subseteq \Lambda$. The lattice $\Lambda$ is called \emph{fine} lattice while its subset $\Lambda^{\prime}$ is called \emph{coarse} lattice. The lattice partition is formed by:
\begin{equation}\label{eq:1d1}
\Lambda / \Lambda^\prime = \{ \boldsymbol{\lambda} + \Lambda^\prime, \; \; \boldsymbol{\lambda} \in \Lambda \}.
\end{equation}

Note that for each $\boldsymbol{\lambda} \in \Lambda$, the set $\boldsymbol{\lambda} + \Lambda^\prime$ is a \emph{coset} of $\Lambda^\prime$ in $\Lambda$. The point $\boldsymbol{\lambda}\;\text{mod}\; \Lambda^\prime$ is called the \emph{coset leader} of $\boldsymbol{\lambda} + \Lambda^\prime$. The number of cosets or the cardinality of $\Lambda / \Lambda^\prime$ is denoted by $\mathcal{M}$ and calculated as:
\begin{equation}\label{eq:1f}
 \mathcal{M} = |\Lambda / \Lambda^\prime| = \text{Vol}(\Lambda^\prime)/\text{Vol}(\Lambda),
\end{equation}
where $\text{Vol}(\Lambda)$ is the volume of the lattice $\Lambda$ and can be calculated as $\text{Vol}(\Lambda) = |\text{det}(\mathbf{G}_{\Lambda})|$. We denote the set of coset leaders by $\Psi = \{\psi_0,\psi_1,\ldots,\psi_{\mathcal{M}-1} \}$.

A \emph{nested lattice code} $\mathcal{L}$ is defined as the set of all coset leaders in the lattice partition $\Lambda / \Lambda^\prime$. In other words, it takes all the lattice points inside the fundamental Voronoi region of the coarse lattice $\Lambda^\prime$:
\begin{equation}\label{eq:1e}
\mathcal{L} = \Lambda \cap \mathcal{V}_{\Lambda^\prime}(\mathbf{0}).
\end{equation}

Due to this geometry property, the fundamental Voronoi region $\mathcal{V}_{\Lambda^\prime}(\mathbf{0})$ is also called the \emph{shaping region}. Shaping is essential in designing practical lattice codes because a finite section of the lattice points must be selected to satisfy a transmission power constraint for a communication system.

Denote the \emph{code rate} of the nested lattice code by $R$. The code rate is measured in bit per dimension and can be calculated as:
\begin{equation}\label{eq:ir}
R = \frac{1}{n}\text{log}_2(\mathcal{M}),
\end{equation}
where $n$ is the dimension of the lattice and $\mathcal{M}$ is the cardinality defined in (\ref{eq:1f}).

We now look at some figures of merit that used to measure the goodness of the lattices. Particularly, we focus on the shaping performance of the lattices. First of all, we define the \emph{second moment} $P(\Lambda)$ as the average energy per dimension of a uniform distribution over the fundamental Voronoi region $\mathcal{V}_{\Lambda}(\mathbf{0})$:
\begin{equation}\label{eq:ik}
P(\Lambda) = \frac{1}{n\text{Vol}(\Lambda)}\int_{\mathcal{V}_{\Lambda}(\mathbf{0})} \|\mathbf{x}\|^2 d\mathbf{x}.
\end{equation}

The \emph{normalised second moment} (NSM) of lattice $\Lambda$ is defined as:
\begin{equation}\label{eq:nsm}
G(\Lambda)=\frac{P(\Lambda)}{\text{Vol}(\Lambda)^{\frac{2}{n}}}.
\end{equation}

The shaping gain $\gamma_s(\Lambda)$ is defined as the energy gain by achieving the reduction of the average energy of a lattice constellation compared with the constellation points that form an $n$-dimensional cube. It can be calculated as:
\begin{equation}\label{eq:sg}
\gamma_s(\Lambda) = \frac{1/12}{G(\Lambda)},
\end{equation}
where $\frac{1}{12}$ is the NSM of an $n$-dimensional cubic lattice which is deemed as the baseline. A lattice with a smaller normalised second moment is always desirable as its shaping gain is higher. When the dimension approaches infinite, there exist a sequence of lattices that can achieves the optimal shaping gain:
\begin{equation}
\lim_{n \rightarrow \infty}\gamma_s(\Lambda_n) = \frac{\pi e}{6}.
\end{equation}

\section{Multi-Dimensional IRA Lattice Codes}
In this section we present the proposed multi-dimensional IRA lattice codes. We consider the channel to be a complex AWGN channel where the input is non-binary, which means asymmetric-output in general. For this channel, different transmitted symbols have different error resistance to the non-binary AWGN noise. Thus the decoding errors for different symbols are different.

\subsection{IRA Lattices Construction}
We begin with the construction of our lattice codes. The lattice codes are constructed via Construction A \cite{Conway99}. The error performance of Construction A lattices heavily depends on the underlying error correction codes. Thus we choose IRA codes as they have been shown to be capacity approaching in AWGN channels and has lower encoding complexity than that of general LDPC codes \cite{Jin00,Chiu10,7008249,7124694,Yang15}.

In this work, we extend the conventional Construction A method to a more generic case which is not merely limited to two-dimensional lattices. Denote a non-binary IRA codes over $\text{GF}(p^M)$ by $\mathcal{C}$, where $p$ is a prime number and $M$ is a positive integer. The IRA encoder takes length $K$ input messages and produces length $N$ codewords. Here $K \leq N$ and all the encoding operations are over $\text{GF}(p^M)$. We denote the Construction A lattice by $\Lambda_{\mathcal{C}}$. It can be generated via:
\begin{equation}\label{eq:31}
\Lambda_{\mathcal{C}} = \{ \boldsymbol{\lambda} = \phi(\mathcal{C})+\xi\mathcal{R}^N \},
\end{equation}
where $\xi \in \mathcal{R}$ and $\mathcal{R}$ is a lattice; $\phi(.)$ is a homomorphism mapping function that maps each codeword component to the elements in the lattice partition:
\begin{equation}\label{eq:31a}
\phi:\mathbb{F}_p^M\rightarrow \mathcal{R}/\xi\mathcal{R}.
\end{equation}
Note that $N$ in (\ref{eq:31}) should be a multiple of $M$ in (\ref{eq:31a}).

It is also noteworthy that in conventional Construction A, $\mathcal{R}$ can be any principal ideal domains (PID) such as rational integers $\mathbb{Z}$ and Gaussian integers $\mathbb{Z}[i]$. In that case, the lattice partition forms a quotient ring that is isomorphic to a finite field. In most cases where $\mathcal{R}$ is a multi-dimensional lattice, the lattice partition forms a quotient group \cite{Oggier13}.

In (\ref{eq:31a}), the $\mathcal{R}$-lattice is partitioned into $p^M$ numbers of cosets where each coset has a coset leader. For designing \emph{finite} constellations, only coset leaders are used in transmission to satisfy the power constraint requirement. Therefore, using (\ref{eq:ir}), the information rate $R$ for this Construction A lattice is
\begin{equation}\label{eq:aa3}
R = \frac{K}{N}\cdot\frac{1}{n}\log_2(p^M),
\end{equation}
where $n$ is the dimension of the $\mathcal{R}$-lattice.

We now present a specific design example of using the $D_4$ lattice via Construction A. According to \cite{Conway99}, the $D_4$ lattice is a four-dimensional lattice which has the highest sphere packing density in the four-dimensional space. It is defined as:
\begin{equation}\label{eq:32}
D_4 = \bigg\{(x_1,x_2,x_3,x_4) \in \mathbb{Z}^4: \sum\nolimits_{i=1}^4 x_i \in 2 \mathbb{Z}\bigg\}.
\end{equation}
It has the generator matrix in the integer lattice form:
\begin{equation}\label{eq:33}
G_{D_4}= \begin{bmatrix}
-1 & -1 & 0 & 0 \\
1 & -1 & 0 & 0 \\
0 & 1 & -1 & 0  \\
0 & 0 & 1 & -1 \\
\end{bmatrix}.
\end{equation}

As explained in Section II, we use the NSM as the goodness to measure the shaping performance of the lattices. By (\ref{eq:nsm}), we calculate the NSM for $D_4$ is about $0.0766$. Then using (\ref{eq:sg}) we can see that $D_4$ can provide a shaping gain about $0.3657$ dB over the four dimensional cubic lattice.

According to \cite{Natarajan15}. the $D_4$ lattice can be identified as \emph{Hurwitz quaternion integers}:
\begin{equation}\label{eq:34}
\mathbb{H} = \bigg\{a+bi+cj+dk |a,b,c,d \in \mathbb{Z} \;\text{or}\; a,b,c,d \in \mathbb{Z} + \frac{1}{2} \bigg\},
\end{equation}
where $\{1,i,j,k\}$ is the basis of the number system for representing Hurwitz integers.
Addition in $\mathbb{H}$ is component wise whereas multiplication is non-commutative and defined based on the following relations:
\begin{equation}\label{eq:35}
i^2 = j^2 = k^2 = ijk = -1.
\end{equation}

Given $A = a+bi+cj+dk$, the norm of $A$ is:
\begin{equation}\label{eq:36}
N(A) = a^2+b^2+c^2+d^2 \in \mathbb{Z}.
\end{equation}

Consider the following example. In (\ref{eq:31a}), if we let $\xi = 1+2i$, then the homomorphism mapping function becomes:
\begin{equation}\label{eq:37}
\phi:\mathbb{F}_5^2 \rightarrow \mathbb{H}/(1+2i)\mathbb{H}.
\end{equation}
Note that this lattice partition can be further expressed as:
\begin{align}
\mathbb{H}/(1+2i)\mathbb{H} &= \boldsymbol{\lambda}/(1+2i)D_4, \;\; (\boldsymbol{\lambda} \in D_4)\nonumber \\
& \overset{(\ref{eq:1c})}= \boldsymbol{\lambda} - Q_{(1+2i)D_4}\left( \boldsymbol{\lambda} \right) \nonumber \\
& \overset{(a)}= \boldsymbol{\lambda} - (1+2i)Q_{D_4}\left( \frac{\boldsymbol{\lambda}}{(1+2i)}\right),
\end{align}
where (a) follows \cite[Eq. (2.43)]{Zamir15}. The multiplication and division here should follow quaternion arithmetic \cite{Smith03}. For the quantizer $Q_{D_4}$, we follow the approach in \cite{Conway82} to develop the quantization algorithm of finding the closest $D_4$ lattice point to an arbitrary point in $\mathbb{R}^4$. The quantization algorithm has a lower computational complexity compared with ML decoding. It is very useful in the scenario where we perform the $D_4$ lattice partitions. The cardinality of this partition can be calculated by using (\ref{eq:36}) as $N(1+2i)^2 = 25$. In this way, the $D_4$ lattice is partitioned into 25 cosets. Even though $\mathbb{H}$ is a PID \cite{Huang17}, we only have the group homomorphism as the multiplication for $\mathbb{H}$ is non-commutative.

Now we compare the mutual information of the $D_4$ lattice with that of a two-dimensional lattice to see the performance gain introduced by the multi-dimensional lattices. In this paper, the two-dimensional square lattice $\mathbb{Z}^2$ is set to be a benchmark for performance comparison. Note that a finite portion of the $\mathbb{Z}^2$ lattice is known as a quadrature amplitude modulation (QAM). The $\mathbb{Z}^2$ lattice can be identified as Gaussian integers $\mathbb{Z}[i] = \{a+bi:a,b \in  \mathbb{Z} \}$. For fair comparison, we partition both lattices in a way such that the information rates for both lattice partitions are the same.

\begin{figure}[ht!]
	\centering
\includegraphics[scale=0.64]{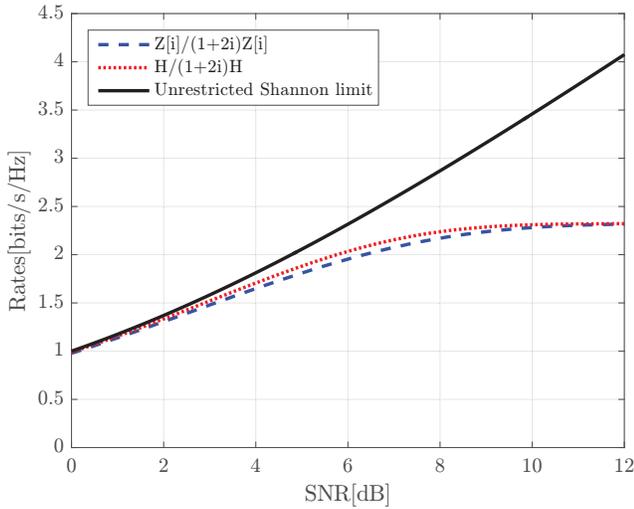}
\caption{Capacities of $\mathbb{H}/(1+2i)\mathbb{H}$ and $\mathbb{Z}[i]/(1+2i)\mathbb{Z}[i]$.}
\label{fig:1}
\end{figure}

We consider the examples of lattice partitions $\mathbb{H}/(1+2i)\mathbb{H}$ and $\mathbb{Z}[i]/(1+2i)\mathbb{Z}[i]$, where both partitions yield the same information rate. This is because using (\ref{eq:ir}) we can obtain the information rates for $D_4$ and $\mathbb{Z}^2$ as $\frac{1}{2}\log_2(25)$ and $\log_2(5)$, respectively. Here the $D_4$ lattice can be deemed as a two-dimensional complex lattice while the $\mathbb{Z}^2$ lattice is a one-dimensional complex lattice. Therefore the dimensions $n$ in (\ref{eq:ir}) for both lattices are 2 and 1, respectively. In other words, the $\mathbb{Z}[i]$ lattice requires one time slot to transmit its lattice point where the $D_4$ lattice requires two time slots to transmit a $D_4$ lattice point.

Given SNR values, the unrestricted Shannon limit for the AWGN channel is plotted in Fig. \ref{fig:1} along with the capacities of the $D_4$ lattice and the $\mathbb{Z}^2$ lattice. As observed from Fig. \ref{fig:1}, the curve for the $D_4$ lattice always lies above that for the $\mathbb{Z}^2$ lattice. Therefore, under the same information rate, we can construct $D_4$ lattice partition based IRA lattice codes that require lower decoding SNR than any IRA lattice codes based on the $\mathbb{Z}^2$ lattice partitions. This is due to the advantage of shaping gain.

\subsection{IRA Lattice Encoder and Its Linearity}

\subsubsection{IRA Lattice Encoder}\label{IRAE}
\begin{figure}[ht!]
	\centering
\includegraphics[scale=0.51]{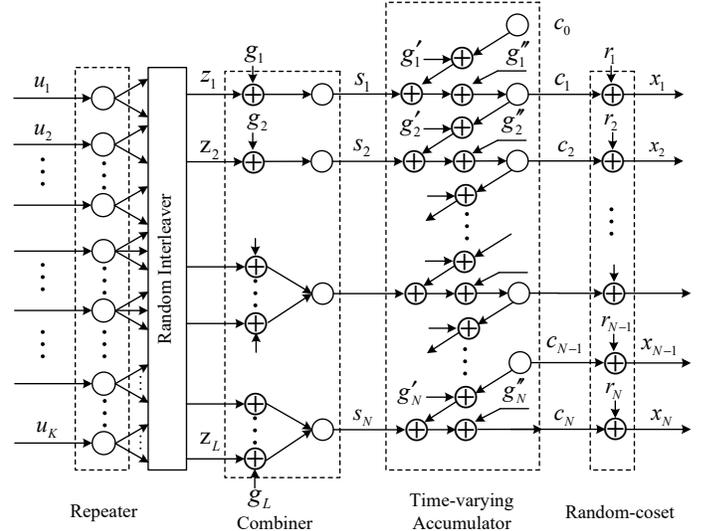}
\caption{Block diagram of the IRA lattice encoder.}
\label{fig:2}
\end{figure}

Here we show our proposed encoder design. The block diagram of the IRA lattice encoder is depicted in Fig. \ref{fig:2}.
First of all, the input to the encoder is a length $K$ message $\mathbf{u}=[u_1,u_2,\ldots,u_K]^T$, where each element $u_k$ for $k = 1,2,\ldots, K$ is taken from the set of coset leaders $\Psi = \{\psi_0, \psi_1, \ldots, \psi_{p^M-1}\}$. This message $\mathbf{u}$ is then fed into a repeater and repeated according to a discrete distribution of $f_1,f_2,\ldots,f_I$, where $f_i \geq 0$ for $i = 1, 2, \ldots ,I$ and $\sum_if_i=1 $. The number $f_i$ represents the fraction of message symbols are repeated by $i$ times. The maximum repeating times is $I$ times, where $I \geq 2$, thus $f_1 = 0$. After repeating, the total number of symbols becomes $L = K\sum_iif_i$.

Next, the repeated symbols are passed into a random interleaver. We denote the interleaved sequence by $\mathbf{z} = [z_1,z_2,\ldots,z_L]^T$. A randomly generated sequence with the same length $\mathbf{g} = [g_1,g_2,\ldots,g_L]^T$ is added to the interleaved sequence $\mathbf{z}$ via $\mathbf{z} \oplus \mathbf{g}$ in an element-wise manner, where ``$\oplus$'' is the modulo-lattice addition defined in (\ref{eq:1d}). Note that each element of $\mathbf{g}$ is randomly and uniformly chosen from the set of coset leaders $\Psi$ such that a linear code constraint is met, which will be introduced later.

The resultant symbols are combined according to a discrete distribution of $b_1,b_2,\ldots,b_J$, where $b_j \geq 0$ for $j = 1,2,\ldots,J$ and $\sum_jb_j=1 $. Here the number $b_j$ represents the fraction of message symbols that are obtained from combining $j$ symbols from the output of the interleaver and the corresponding $j$ addition factors in $\mathbf{g}$. After combining, the message sequence becomes a length $N$ sequence denoted by $\mathbf{s} = [s_1,s_2,\ldots,s_N]^T$, where $N = L\sum_j j b_j$. For $n = 1,...,N$, each symbol $s_n$ is calculated as:
\begin{equation}\label{eq:38}
s_n = (z_{a_n} \oplus g_{a_n}) \oplus \ldots \oplus (z_{a_n+j_n-1} \oplus g_{a_n+j_n-1}),
\end{equation}
where $z_{a_n}$ and $z_{a_n+j_n-1}$ represent the first and last interleaved symbols input to the $n$-th combiner, respectively; $g_{a_n}$ and $g_{a_n+j_n-1}$ are the addition factors with respect to $z_{a_n}$ and $z_{a_n+j_n-1}$; $j_n \in \{1,2,\ldots,J \}$ represents the number of symbols to be combined at the $n$-th combiner; $a_n$ is the index of the first interleaved symbol input to the $n$-th combiner. Note that the combiner is to combine the interleaved messages in order to satisfy the code rate requirement.

The combined message sequence $\mathbf{s}$ is passed into a time-varying accumulator which features a time-varying transfer function determined by two randomly generated lattice sequences $\mathbf{g^\prime} = [g^\prime_1,g^\prime_2,\ldots,g^\prime_N]^T$ and $\mathbf{g^{\prime\prime}} = [g^{\prime\prime}_1,g^{\prime\prime}_2,\ldots,g^{\prime\prime}_N]^T$. All the elements in both sequences are uniformly distributed over the set of coset leaders $\Psi$ such that a linear code constraint is met, which will be introduced later. The output message of the time-varying accumulator is denoted by $\mathbf{c} = [c_1,c_2,\ldots,c_N]^T$. The $n$-th symbol $c_n$, where $n = 1,2,\ldots,N$, is generated by
\begin{equation}\label{eq:39}
c_n = (s_n \oplus (c_{n-1} \oplus g^\prime_n)) \oplus g^{\prime\prime}_n,
\end{equation}
where the initial condition is given as $c_0 = 0$. Here $c_0$ is a dummy parity that is fixed to $0$ and will not be transmitted. It is also noteworthy that the random vectors $\mathbf{g}$, $\mathbf{g}^\prime$ and $\mathbf{g}^{\prime\prime}$ in the encoding structure introduce and realize the permutation-invariance property on \emph{all edges} of a Tanner graph as shown in Fig. \ref{fig:tg} and will be discussed in Section \ref{msi}.

Finally, the output of the accumulator $\mathbf{c}$ adds a random-coset vector $\mathbf{r}$ with length $N$ and become the coded lattice sequence $\mathbf{x}$:
\begin{equation}\label{eq:40}
\mathbf{x} = \mathbf{c} \oplus \mathbf{r}.
\end{equation}
Elements of $\mathbf{r}$ are uniformly distributed over the set of coset leaders $\Psi$. Before transmission, the average energy of codeword symbols is normalised to 1.

Note that although the four lattice sequences $\mathbf{g}$, $\mathbf{g^{\prime}}$, $\mathbf{g^{\prime\prime}}$ and $\mathbf{r}$ are random, they are assumed to be known at both transmitters and receivers prior to transmission. Furthermore, the underlying linear codes for our Construction A lattices can be either systematic or nonsystematic non-binary IRA codes.

\subsubsection{The Linearity of IRA Lattice Codes}
It can be noticed that our proposed lattice encoding structure is different from previous designs. More specifically, instead of using the modulo-lattice multiplication between encoder messages and random lattice sequences in \cite{Qiu16}, we use a different approach by introducing the ``$\oplus$'' operation in the encoding process. However, this difference introduced non-linearity to our codes if $\mathbf{g}$, $\mathbf{g^{\prime}}$ and $\mathbf{g^{\prime\prime}}$ are totally independent, which is not appealing for low complexity decoding. To address this issue, we introduce a constraint on these random sequences to ensure the codes are linear.
\begin{Pro}\label{linear}
The multi-dimensional IRA lattice codes are linear if the $n$-th output element from the encoder satisfies the following conditions:
\begin{equation}\label{eq:LC}
g_{a_n} \oplus \ldots \oplus g_{a_n+j_n-1} \oplus g_n^{\prime} \oplus g_n^{\prime\prime}  = 0.
\end{equation}
\end{Pro}
\begin{IEEEproof}
See Appendix \ref{appendix:1}.
\end{IEEEproof}
Note that this equation has $j_n+2$ elements. We randomly choose any $j_n+1$ elements out of these $j_n+2$ elements to be random and uniformly distributed over the set of coset leaders $\Psi$. The last element is then determined by Eq. (\ref{eq:LC}). One can also notice that the linearity condition excludes the random-coset vector $\mathbf{r}$. This is because the random-coset vector is independent of the encoder's messages and is always removed before decoding. If the random-coset vector is included in the condition, the output-symmetric effect in the non-binary AWGN channel will vanish.

\subsection{Tanner Graph}
\begin{figure}[ht!]
	\centering
\includegraphics[scale=0.56]{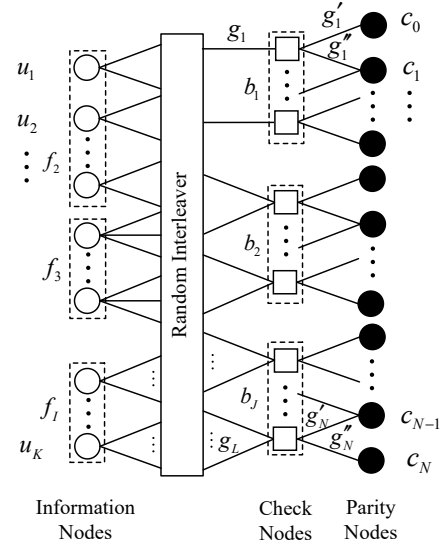}
\caption{Tanner graph of the IRA lattice codes.}
\label{fig:tg}
\end{figure}

Similar to conventional binary IRA codes in \cite{Jin00}, our multi-dimensional IRA lattice codes can be represented by a Tanner graph as shown in Fig. \ref{fig:tg}.

The Tanner graph is a bipartite graph with variable nodes and check nodes. In the figure, variable nodes are represented by circles while check nodes are represented by squares. There are $N+K$ variable nodes on the Tanner graph. The $K$ variable nodes that placed on the left, are called \emph{information nodes}. They represent the $K$ repeaters in the encoder. The degree distribution of information nodes with degree $i$ is denoted by $f_i$ in the figure. This means that the fraction of information nodes are connected to $i$ check nodes. Note that the random interleaver here introduces randomness in the edges between information nodes and check nodes. This randomness can prevent short cycles in the Tanner graph which leads to a better decoding performance \cite{Johnson05}. On the right of the Tanner graph, there are $N$ variable nodes which are called \emph{parity nodes}, representing the output $\mathbf{c}$ from the time-vary accumulator. In the middle of the Tanner graph, there are $N$ check nodes, representing $N$ combiners. The degree distribution of check nodes with degree $j+2$ is denoted by $b_j$ which represents the fraction of check nodes connected to $j$ information nodes and 2 parity nodes. Note that the random-coset vector $\mathbf{r}$ is removed before performing decoding, thus it is not shown in the Tanner graph.

Now consider the $n$-th check node with degree $j+2$, according to (\ref{eq:38}), (\ref{eq:39}) and the Tanner graph in Fig. \ref{fig:tg}, the parity-check equation at the $n$-th check node is
\begin{align}\label{eq:41}
&(z_{a_n} \oplus g_{a_n}) \oplus \cdots \oplus (z_{a_n+j_n-1} \oplus g_{a_n+j_n-1})\oplus \nonumber \\
& (c_{n-1} \oplus g^\prime_n) \oplus ( c_n^{-1}  \oplus g^{\prime\prime}_n) = 0,
\end{align}
where $c_n^{-1} \oplus c_n = 0$. Note that in the Tanner graph, $c_0$ is a dummy bit and will not be transmitted.

We decompose the elements on the left hand side of Equation (\ref{eq:41}) into two vectors:
\begin{equation}\label{eq:42}
\mathbf{t}_n = [z_{a_n}, \ldots ,z_{a_n+j_n-1}, c_{n-1}, c_n^{-1}].
\end{equation}
\begin{equation}\label{eq:43}
\mathbf{h}_n = [g_{a_n}, \ldots ,g_{a_n+j_n-1}, g^\prime_n, g^{\prime\prime}_n] .
\end{equation}
The first vector $\mathbf{t}_n$ represents the symbols coming from the variable nodes connected to the $n$-th check node. More specifically, $z_{a_n}, \ldots ,z_{a_n+j_n-1}$ are from information nodes while $c_{n-1}$ and $c_n^{-1}$ are from parity nodes. The second vector $\mathbf{h}_n$ represents the addition factors on the corresponding edges of the $n$-th check nodes as shown in Fig. \ref{fig:tg}.

\subsection{IRA Lattice Decoder}
As shown in the previous section, the multi-dimensional IRA lattice codes have a Tanner graph representation. Therefore, we can employ a modified belief prorogation (BP) decoding algorithm to decode our lattice codes.

The decoder attempts to recover the source message $\mathbf{u}$ from the noisy observation of the AWGN channel output $\mathbf{y} = \mathbf{x}+\mathbf{n}_z $, where $\mathbf{n}_z \thicksim \mathcal{CN}(0,\sigma_{ch}^2) $ denotes the complex AWGN noise. Before decoding, we first need to calculate the symbol-wise a posterior probability (APP) of each coset leader and for each lattice codeword component $x_n$, which is written as:
\begin{equation}\label{eq:app1}
P(x_n|y_n) = \frac{p(y_n|x_n)p(x_n)}{p(y_n)}, \,\, \text{for}\, \, n = 1,2,\ldots,N.
\end{equation}

For the sake of simplicity, We let
\begin{equation}\label{eq:app3}
P_{\psi_k}[n] = P(x_n = \psi_k|y_n),
\end{equation}
where $k = 0,1,\ldots,p^M-1$ and $\psi_k$ is the $k$-th coset leader. Since the transmitted codeword symbol is $x_n = c_n+r_n$, where $r_n$ is uniformly distributed over $\Psi$, thus the distribution for $x_n$ is also uniform over $\Psi$. Therefore, Eq. (\ref{eq:app3}) can be written as
\begin{equation}\label{eq:app5}
P_{\psi_k}[n] = \frac{P(y_n|x_n = \psi_k)}{\sum_{k=0}^{p^M-1} P(y_n|x_n = \psi_k)},
\end{equation}
and
\begin{equation}\label{eq:app4}
P(y_n|x_n = \psi_k) = \frac{1}{\sqrt{2\pi\sigma_{ch}^2}}\exp\left(-\frac{\|y_n-\sqrt{\text{SNR}}\psi_k\|^2}{2\sigma_{ch}^2}\right),
\end{equation}
In this way, we have $\sum_{k=0}^{p^M-1}P_{\psi_k}[n]=1$.

In (\ref{eq:app4}), $\psi_k$ and $y_n$ both are vectors with length equal to the dimension of the lattice. In our design example, $\psi_k$ is a $D_4$ lattice point with four dimensions. We perform the symbol-wise maximum-likelihood detection. Considering that practical systems can only transmit and receive one two-dimensional signal at each time slot, the detection is a joint detection for two two-dimensional signals.

We denote the APP vector by $\mathbf{P}[n]$ where
\begin{equation}\label{eq:app6}
\mathbf{P}[n] = [P_{\psi_0}[n],P_{\psi_1}[n],\ldots,P_{\psi_{p^M-1}}[n]]^T.
\end{equation}

Then the above APP vectors are fed into a coset remover to obtain the APP vectors with respect to $\mathbf{c}$ in (\ref{eq:40}) as the message before adding the random-coset vector $\mathbf{r}$. We denote the APP vector after removing coset by $\mathbf{P}^\prime[n]$:
\begin{align}\label{eq:app2}
\mathbf{P}^\prime[n] & = \mathbf{P}(c_n|y_n) \nonumber \\
&= [P_{\psi_0 \ominus r_n}[n],P_{\psi_1  \ominus r_n}[n],\ldots,P_{\psi_{p^M-1}  \ominus r_n}[n]]^T.
\end{align}
where $\ominus$ is defined in (\ref{eq:1dss}). The resultant APP vector $\mathbf{P}^\prime[n]$ is then passed into a BP decoder.

The decoder updates the information between check nodes and variable nodes in an iterative manner. We denote the message from the $m$-th variable node to the $n$-th check node by $\bm{r}(m,n)$. The message passed from the $n$-th check node to the $m$-th variable node is denoted by $\bm{l}(n,m)$. Both vectors are probability vectors with dimension $p^M$. Use the Tanner graph in Fig. \ref{fig:tg}, we let $\mathcal{A}(m)$ and $\mathcal{B}(n)$ represent the set of check nodes connected to the $m$-th variable node and the set of variable nodes adjacent to the $n$-th check node, respectively. Without the loss of generality, let the index of information nodes be from $1$ to $K$ and the index of parity nodes be from $(K+1)$ to $(K+N)$ of the variable nodes. The decoding steps can be summarized in the following.

\subsubsection{Initialisation step}
According to the Tanner graph in Fig. \ref{fig:tg}, the channel output must go through the parity nodes first. Thus for all edges $(m,n)$ between the parity nodes and the check nodes in the Tanner graph, the initial message $\bm{r}(m,n)$ is the channel APP in (\ref{eq:app2}):
\begin{align}\label{eq:de1}
\bm{r}(m,n) = \mathbf{P}^\prime[m-K], \,\,\,\,\text{for}\,\, &m = K+1,\ldots,K+N \nonumber \\
&n = 1,2,\ldots,N.
\end{align}

For all edges $(m,n)$ between the information nodes and the check nodes in the Tanner graph, we let
\begin{align}\label{eq:de1a}
r_k(m,n) = \frac{1}{p^M}, \,\,\,\, \text{for}\,\, &k =0,1,\ldots,p^M-1 \nonumber \\
&m = 1,2,\ldots,K.
\end{align}


\subsubsection{Update the check nodes to variable nodes messages}
For all edges $(m,n)$ that connected to the $n$-th check node, generate the probability vector $\bm{l}(n,m)$ with its $k$-th element given by
\begin{equation}\label{eq:de2}
l_k(n,m) = \sum_{\substack{t_1,\ldots,t_{j_n-1} \in \Psi \\ \bigoplus_{i=1}^{j_n-1}(t_i \oplus h_i) \oplus \psi_k \oplus (h_{j_n})=0}}\prod\nolimits_{i=1}^{j_n-1}r_{t_i}^{(i)},
\end{equation}
where $\bigoplus$ is the summation performed by $\oplus$; $j_n$ is the degree of the $n$-th check node; $\bm{r}^{(1)},\ldots,\bm{r}^{(j_n-1)}$ are the incoming messages from all the connected variable nodes except the $m$-th variable node, i.e., $\{\bm{r}(m^\prime,n):m^\prime \in \mathcal{B}(n) \setminus m \}$; $t_1,\ldots,t_{j_n-1}$ are the lattice symbols from the associated variable nodes; $h_{j_1},h_{j_2},\ldots,h_{j_n-1}$ are the addition factors on the corresponding edges and $h_{j_n}$ denotes the addition factor for the edge $(m,n)$. Note that the calculations of the check node messages are different from that in conventional IRA decoding as the parity-check equations and the associated arithmetic are different.

\subsubsection{Update the variable nodes to check nodes messages}
For all edges $(m,n)$ between the variable nodes and the check nodes in the Tanner graph, generate the probability vector $\bm{r}(n,m)$ with the $k$-th element given by
\begin{equation}\label{eq:de3}
r_k(m,n) = \frac{\gamma^{(n)}_k\prod_{i=1}^{j_m-1}l_k^{(i)}}{\sum_{k^\prime=0}^{p^M-1} \gamma^{(n)}_{k^\prime}\prod_{i=1}^{j_m-1}l_{k^\prime}^{(i)}},
\end{equation}
where $j_m$ denotes the degree of the $m$-th variable node; $\bm{l}^{(1)},\ldots,\bm{l}^{(j_m-1)}$ denote the incoming messages from all the connected check nodes except the $n$-th check node, i.e., $\{\bm{l}(n^\prime,m):n^\prime \in \mathcal{A}(m)/n \}$; $\gamma^{(n)}_k = r_k(m,n)$ in (\ref{eq:de1}) for $m = K+1,\ldots,K+N$ when the messages are from parity nodes to the $n$-th check node and $\gamma^{(n)}_k = r_k(m,n)$ in (\ref{eq:de1a}) for $m = 1,\ldots,K$ when the messages are from information nodes to the $n$-th check node.

\subsubsection{Stopping condition}
For each iteration, make the hard decision on the $m$-th variable node by calculating
\begin{equation}\label{eq:de4}
\hat{\delta}_n = \argmax_k\frac{\gamma^{(n)}_k\prod_{i=1}^{j_m}l_k^{(i)}}{\sum_{k^\prime=0}^{p^M-1} \gamma^{(n)}_{k^\prime}\prod_{i=1}^{j_m}l_{k^\prime}^{(i)}},
\end{equation}
for $n = 1,2,\ldots,K+N$. It contains information from all the connected edges. If the hard decision results  $\hat{\delta}_1,\hat{\delta}_2,\ldots,\hat{\delta}_{K+N}$ satisfy the parity-check equations in (\ref{eq:41}) or a predetermined maximum number of iterations is reached, then stop; otherwise go to Step 2).

The calculation in (\ref{eq:de2}) has a very high computational complexity if the cardinality of the lattice partition $p^M$ is very large. We follow \cite{Richardson01} to employ DFT and IDFT in our lattice decoding process to reduce the complexity.

First we need to introduce some important notations which will be used in the rest of the paper. Define a probability vector as $\boldsymbol{\rho} = [\rho_{\psi_0}, \rho_{\psi_1},\ldots,\rho_{\psi_{p^M-1}}]$ representing the probability of a lattice point being $\psi_0, \psi_1, \ldots, \psi_{p^M-1}$. In addition, the probability vector must satisfy $\rho_{\psi_k} \geq 0$ and $\sum_{k=0}^{p^M-1}\rho_{\psi_k} = 1$. Given a probability vector $\boldsymbol{\rho}$ and $\chi \in \Psi$, we define the $\oplus \chi$ operation as the following
\begin{equation}\label{eq:de5}
 \boldsymbol{\rho}^{\oplus\chi}=[\rho_{\psi_0 \oplus \chi}, \rho_{\psi_1\oplus \chi},\ldots,\rho_{\psi_{p^M-1}\oplus \chi}].
\end{equation}

Now consider the expression in (\ref{eq:de2}), an equivalent expression can be written as
\begin{equation}\label{eq:de6}
\bm{l} = \bigg[\bigotimes\nolimits_{i=1}^{j_n-1}\left(\bm{r}^{(i)}\right)^{\ominus h_i} \bigg]^{\ominus h_{j_n}},
\end{equation}
where $\bm{l}$ is the vector that contains elements $l_k$, $k = 0,1,\cdots, p^M-1$ in (\ref{eq:de2}) and the ``$\bigotimes$'' operator performs the modulo-lattice convolution between two vectors. It produces a vector whose $k$-th component is:
\begin{equation}\label{eq:de7}
[\bm{r}^{(1)} \otimes \bm{r}^{(2)}]_k = \sum_{\chi \in \Psi} r_{\chi}^{(1)}\cdot r_{\psi_k \ominus \chi}^{(2)}, \,\, \text{for}\, k =0,1,\ldots,p^M-1.
\end{equation}
This convolution can be evaluated by using $M$-dimensional DFT and IDFT \cite{Dudgeon84}. In this way, (\ref{eq:de6}) can be evaluated as
\begin{equation}\label{eq:de8}
\bm{l} = \Bigg[\text{IDFT} \bigg[\prod\nolimits_{i=1}^{j_n-1}\text{DFT}\left( \left( \bm{r}^{(i)}\right)^{\ominus h_i}\right)  \bigg]\Bigg]^{\oplus h_{j_n}},
\end{equation}
where the multiplication of the DFT vectors is performed in a component-wise manner. A further reduction in complexity of implementation can be obtained by using fast Fourier transform and inverse fast Fourier transform algorithms.

\subsection{Complexity of IRA lattice codes}
In this section, the complexity of our multi-dimensional IRA lattice codes will be investigated and compared to that of the IRA lattice codes with two-dimensional lattice partitions.Note that both lattice codes are built from Construction A. The underlying linear code for our design is over $\mathbb{F}_p^2$ while the linear codes for the design with two-dimensional lattices is over $\mathbb{F}_p$ \cite{Qiu16}.

First, we focus on the complexity of symbol-wise detection. For an ML detector, the detection is based on the entire constellation. Thus, for a two-dimensional constellation with size $q$, the computational complexity is in the order of $O(q)$. In our design, we have a four-dimensional constellation with size $q^2$, the computational complexity is $O(2q^2)$. The ``2'' here is due to the joint detection for two two-dimensional symbols. The computational complexity of the nonbinary BP decoding is in the order of $O(q\log_2 q)$ when FFT is employed for check node calculations \cite{Ganepola08}. For our decoder to decode lattice codes with four-dimensional lattice partitions, the complexity is $O(q^2\log_2 q^2)$. Compared with our previous coding scheme with two-dimensional lattice partitions, the complexity of the code design in this work is $2q$ times higher.

For Construction A lattices, it has been shown in \cite{Zamir04} that the finite field size of the underlying linear code has to be large enough to achieve the capacity. Therefore, we have traded the complexity to attain better performance by introducing multi-dimensional lattice partitioned in our design.

\section{Design and Analysis of Multi-dimensional IRA Lattice Codes}
In this paper, the analysis of our multi-dimensional IRA lattice codes focus on the average behaviour of randomly selected codes from an ensemble of codes. First, let $\alpha_i$ be the fraction of interleaver's edges that connected to the information nodes with degree $i$ and let $\beta_j$ be the fraction of interleaver's edges that are connected to the check nodes with degree $j+2$. Recall in Section III-B that $i = 2,3,\ldots,I$ and $j = 1,2,\ldots,J$. The additional ``2'' here means every check node has two deterministic connections from the connected parity nodes as shown in Fig. \ref{fig:tg}. Following \cite{Jin00}, the edge degree distributions of our multi-dimensional IRA lattice codes can be written as
\begin{equation}\label{eq:lam}
\alpha(x) = \sum\nolimits_{i=2}^I\alpha_ix^{i-1}.
\end{equation}
\begin{equation}\label{eq:rho}
\beta(x) = \sum\nolimits_{j=1}^J\beta_jx^{j-1}.
\end{equation}
Given $\alpha$, $\beta$, the type of lattice $\mathcal{R}$ and the scaling factor $\xi$ in (\ref{eq:31a}), we define an $(\alpha, \beta, \xi,\mathcal{R})$ ensemble as the
set of our multi-dimensional IRA lattice codes obtained via Construction A.

\subsection{Modeling the Decoder's Message Distributions}\label{msi}
In our multi-dimensional IRA lattice codes, the soft information propagated in the iterative decoder can be modeled by a multi-dimensional LLR vector. Even though APP is used in our iterative decoder, it is common to use LLR in EXIT chart analysis. Note that APP and LLR are different but equivalent representations of the decoder's soft information. In order to track the convergence behaviour of the iterative decoding, multi-dimensional EXIT charts may be required. However, developing these EXIT chart functions can be very difficult. To deal with this challenge, the new encoding structure is proposed. We will prove that using this structure, the densities of the messages in BP decoder can attain permutation-invariance and symmetry properties. With these two properties, the densities of the decoder's messages can be represented as a single parameter. In this way, our method only needs to track one-dimensional variables rather than the true densities of the multi-dimensional LLR vectors. In addition, the symmetry property enables to use all-zero lattice codeword assumption in the \emph{EXIT chart analysis}. As such, the expression of mutual information in the EXIT chart analysis can be simplified.

We first introduce some useful definitions and notations in the following.

\subsubsection{Preliminaries}
Following the definition in \cite{Li03}, we define the LLR values for a given probability vector $\boldsymbol{\rho}$ as
\begin{equation}\label{eq:LLR}
\omega_{\psi_k} = \log \left(\frac{\rho_{\psi_0}}{\rho_{\psi_k}} \right), \,\text{for}\, k = 0,1,\ldots,p^M-1.
\end{equation}
It is intuitive that $\omega_{\psi_0} = 0$.

The $p^M$-dimensional LLR vector is then defined as $\boldsymbol{\omega} = [\omega_{\psi_0}, \omega_{\psi_1},\ldots,\omega_{\psi_{p^M-1}}]^T$. Note that unlike most LLR definitions, we include the element $\omega_{\psi_0}$ in the LLR vectors as it is associated with our analysis of permutation-invariance which will be introduced shortly. When we apply the $\oplus \chi$ operation defined in (\ref{eq:de5}) on the LLR value $\omega_{\psi_k}$, we have:
\begin{equation}\label{eq:LLR1}
\omega_{\psi_k}^{\oplus \chi} = \log \left(\frac{\rho_{\psi_0 \oplus \chi}}{\rho_{\psi_k \oplus \chi}} \right) = \omega_{\psi_k \oplus \chi} - \omega_{\psi_0 \oplus \chi}.
\end{equation}

A $p^M$-dimensional probability-vector random variable is defined as $\mathbf{P} = [P_{\psi_0},P_{\psi_1},\ldots,P_{\psi_{p^M-1}}]^T$ that only takes valid probability values. The associated $p^M$-dimensional LLR-vector random variable is defined as $\mathbf{W} = [W_{\psi_0},W_{\psi_1},\ldots,W_{\psi_{p^M-1}}]^T$.

Now we introduce the definitions of the symmetry and permutation-invariance properties and explain how we can achieve these properties.

\subsubsection{Symmetry}
Recall in Section III-B, we add a random-coset vector $\mathbf{r}$ at the end of the encoder. The random-coset elements are randomly chosen and uniformly distributed over the set of coset leaders $\Psi$. Thus we have the following theorem.

\begin{The}\label{OS}
Adding a random-coset vector $\mathbf{r}$ to the encoder output $\mathbf{c}$, where $\mathbf{r}$ is uniformly distributed over $\Psi$, can produce the output-symmetric effect in non-binary input AWGN channels.
\end{The}

\begin{IEEEproof}
See Appendix \ref{appendix:4}.
 \end{IEEEproof}
Similar to the non-binary LDPC codes in \cite{Bennatan06}, the LLR random vectors are symmetric under the output-symmetric effect. The symmetry property of an LLR random vector is defined as follows.

\begin{Def}
Given an LLR random vector $\mathbf{W}$ and an $r \in \Psi$, $\mathbf{W}$ is symmetric if and only if $\mathbf{W}$ satisfies
\begin{equation}\label{eq:SS}
\text{Pr}[\mathbf{W} = \boldsymbol{\omega} ] = e^{\omega_{\psi_k}}\text{Pr}[\mathbf{W} = \boldsymbol{\omega}^{\oplus r}]
\end{equation}
for all LLR vectors $\boldsymbol{\omega}$ and all $r \in \Psi$.
\end{Def}

With this property, the probability of decoding error is equal for any transmitted codeword \cite{Bennatan06}. In other words, the symmetry property removes the dependence of the decoder's LLRs on transmitted codewords \cite{Richardson01}. Therefore, we can use all-zero lattice codewords in our EXIT chart analysis.

\subsubsection{Permutation-Invariance}\label{PI}
We start with the definition of permutation-invariance \cite[Section 2.6]{Severini05} on a probability-vector random variable. Then we will show that our approach can achieve this property under our proposed structure.

\begin{Def}
A probability-vector random variable $\mathbf{X} = [X_0,X_1,X_2 \ldots]$ is permutation-invariant if for any permutation $\varpi$ of the indices such that the random vector $\varpi(\mathbf{X}) = [X_{\varpi(0)},X_{\varpi(1)},X_{\varpi(2)},\ldots]$ is distributed identically with $\mathbf{X}$.
\end{Def}

Under this property, all the random variables in $\mathbf{X}$ are identically distributed (but may not be independent). Therefore, changing the order of the elements in $\mathbf{X}$ will not change the distribution of $\mathbf{X}$ .

Recall in Section III-B, our codes have three randomly generated sequences added to the encoder's messages. This leads to a symbol level permutation (the permutation from a coset leader to another coset leader) on the messages. The densities of these messages can be shown to have the permutation-invariance property. Now, we have the following theorem:

\begin{The}\label{PERI}
Given a $p^M$-dimensional probability-vector random variable $\mathbf{P}$ and a $\chi \in \Psi$, the random vector $\mathbf{P}^{\oplus \chi} = [P_{\psi_0 \oplus \chi}, P_{\psi_1\oplus \chi},\ldots,P_{\psi_{p^M-1}\oplus \chi}]$ is identically distributed with $\mathbf{P}$. Therefore $\mathbf{P}$ is permutation-invariant.
\end{The}
\begin{IEEEproof}
See Appendix \ref{appendix:2}.
\end{IEEEproof}
This theorem can be carried over straightforwardly to LLR representation. Thus we have the following lemma:

\begin{Lem}\label{PERIW}
Let $\mathbf{W}= [W_{\psi_0},W_{\psi_1},\ldots,W_{\psi_{p^M-1}}]^T$ be an LLR-vector random variable such that $W_{\psi_k} = \log \left(\frac{P_{\psi_0}}{P_{\psi_k}} \right), \,\text{for}\, k = 0,1,\ldots,p^M-1$. If $\mathbf{P}$ is permutation-invariant, then $\mathbf{W}$ is also permutation-invariant.
\end{Lem}
\begin{IEEEproof}
See Appendix \ref{appendix:3}.
\end{IEEEproof}
Therefore, under the BP decoding, the messages passed within the Tanner graph of our codes satisfy all the symmetry and permutation-invariance properties.

\subsubsection{Gaussian Approximation}
With the symmetry and permutation-invariance properties, the $p^M$-dimensional LLR can be modeled using a multivariate Gaussian distribution \cite{Bennatan06}:
\begin{equation}\label{eq:ga}
f_{\mathbf{W}}(\boldsymbol{\omega})=\frac{1}{(2\pi)^{\frac{p^M}{2}}|\boldsymbol{\Sigma}|^{\frac{1}{2}}}\exp\left(-\frac{1}{2}(\boldsymbol{\omega}-\mathbf{m})^T\boldsymbol{\Sigma}^{-1}(\boldsymbol{\omega}-\mathbf{m})\right),
\end{equation}
with mean vector $\mathbf{m}$ and covariance matrix $\boldsymbol{\Sigma}$ given by
\begin{equation}\label{eq:ga1}
\mathbf{m}=
\begin{bmatrix}
    \frac{\sigma^2}{2}       \\
     \frac{\sigma^2}{2}        \\
    \vdots \\
     \frac{\sigma^2}{2}
\end{bmatrix}
\,\text{and}\,\,\boldsymbol{\Sigma}=
\begin{bmatrix}
    \sigma^2 & \frac{\sigma^2}{2} & \cdots & \frac{\sigma^2}{2}   \\
    \frac{\sigma^2}{2} & \sigma^2 & \cdots & \frac{\sigma^2}{2}   \\
    \vdots & \vdots & \ddots & \vdots  \\
    \frac{\sigma^2}{2} & \cdots & \cdots & \sigma^2
\end{bmatrix}.
\end{equation}
More specifically, $m_i = \frac{\sigma^2}{2}$ for $i = 1,2,\ldots,p^M$, and $\boldsymbol{\Sigma}_{i,j} = \sigma^2$ if $i=j$ and $\frac{\sigma^2}{2}$ otherwise. As a result, the density of the $p^M$-dimensional LLR is completely described by a single parameter $\sigma$. It is worth mentioning that our definition of LLR random vector is $p^M$-dimensional rather than $p^M-1$ in the literature. This is because the $\oplus \chi$ operation will change the position of $W_{\psi_0}$. Thus we need to use a $p^M$-variate Gaussian distribution to model the $p^M$-dimensional LLR.

\subsection{Convergence Analysis}
EXIT charts track the mutual information between the transmit lattice symbol $u$ and the LLR random vector $\mathbf{W}$. With the all-zero lattice codeword assumption, the mutual information can be evaluated according to \cite{Bennatan06}
\begin{equation}\label{eq:exit1}
I(u;\mathbf{W}) = 1-E\Bigg[\log_{p^M}\left(\sum\nolimits_{i=0}^{p^M-1}e^{-w_i} \right) \Bigg|u = 0 \Bigg],
\end{equation}
where $\mathbf{W}$ is modeled by (\ref{eq:ga}) and (\ref{eq:ga1}). Thus, the mutual information is a function of the single parameter $\sigma$. For simplicity, we let $J(\sigma) = I(u;\mathbf{W})$ as every value of $\sigma$ corresponds to a value of $I(u;\mathbf{W})$. Since the mapping is bijective, we can also define the inverse function $J(.)^{-1}$ to obtain $\sigma$ when given $I(u;\mathbf{W})$.

In the EXIT chart analysis, variable nodes are treated as a component decoder while the combiners and the time-varying accumulator together is treated as another decoder. As such, we compute the variable-node decoder (VND) curve and the check-node decoder (CND) curve. The argument of each curve is denoted as $I_A$ and the value of the curve is denoted as $I_E$, representing a priori input and the extrinsic output of each component decoder. The details of obtaining the transfer functions will be explained next.
\subsubsection{EXIT Function for VND}
For a variable node with $i_m$ degrees, the output mutual information of the VND for this type of variable nodes is given by \cite{Brink03}:
\begin{equation}\label{eq:VND}
I_{E,VND}(I_A,i_m)\approx J\left( \sqrt{(i_m-1)}J^{-1}(I_A)  \right).
\end{equation}
For a given VN degree distribution $(i,\alpha_i)$, the EXIT function for the VND of the entire IRA code is:
\begin{equation}\label{eq:VND1}
I_{E,VND}(I_A) =  \sum\nolimits_{i=2}^I \alpha_iI_{E,VND}(I_A;i).
\end{equation}

\subsubsection{EXIT Function for CND}
For a check node with degree $j_n$, we use a numerical method to obtain the approximated EXIT functions as there is no closed-form expression in the literature.

For a given $I_A$, we obtain the corresponding parameter using $\sigma = J^{-1}(I_A)$. Then the input a priori LLR vectors are generated according to (\ref{eq:ga}) and (\ref{eq:ga1}). For a given SNR, generate the all-zero lattice codeword, three random sequences $\mathbf{g}$, $\mathbf{g^\prime}$, $\mathbf{g^{\prime\prime}}$, a random-coset vector $\mathbf{r}$ and an AWGN channel noise sequence with variance of $\sigma_{ch}^2$. We calculate the channel APPs by following (\ref{eq:app1}) to (\ref{eq:app2}) and then substitute the results into (\ref{eq:LLR}) to obtain the channel input LLR $\mathbf{W}_{ch}$. Given $\mathbf{g}$, $\mathbf{g^\prime}$, $\mathbf{g^{\prime\prime}}$, $\mathbf{r}$, $j_n$ and $\mathbf{W}_{ch}$, we perform BP decoding with one iteration to produce the output LLR. The $I_{E,CND}(I_A)$ associated with the check node degree $j_n$ is obtained by substituting the output LLR into (\ref{eq:exit1}).

For a given CN degree distribution $(j,\beta_j)$, the EXIT function for the CND of the entire IRA code can be obtained by:
\begin{equation}\label{eq:CND}
I_{E,CND}(I_A,\sigma_{ch}) =  \sum\nolimits_{j=1}^J \beta_jI_{E,CND}(I_A;j,\sigma_{ch}).
\end{equation}

\subsection{Design Examples}
Based on our EXIT functions, we now employ the EXIT chart curve fitting technique \cite{Brink03} to find the optimal CN and VN degree distributions such that the area between the CN curve and the VN curve is minimized. First, we carefully select an appropriate CN degree distribution. Then, we fit the EXIT curve of VND to CND by using linear programming to optimize the degree distribution for VN. Next, we update the CN degree distribution based on the optimized VN degree distribution. The optimization for the degree distribution of CN and VN are carried out in an iterative manner. Note that we have set the minimum gap between the VND curve and the CND curve to be greater than zero but not too large, e.g., 0.0001. In this way, the produced VND curve do not intersect with the CND curve and both curves create a narrow tunnel. The number of optimization iteration is set to 10 as more iterations does not improve the optimization results further.

An example of an EXIT chart for our multi-dimensional IRA lattice codes over $\mathbb{H}/(1+2i)\mathbb{H}$ with code rate of $\frac{2}{3}$ is illustrated in Fig. \ref{fig:6}.
\begin{figure}[ht!]
	\centering
\includegraphics[scale=0.64]{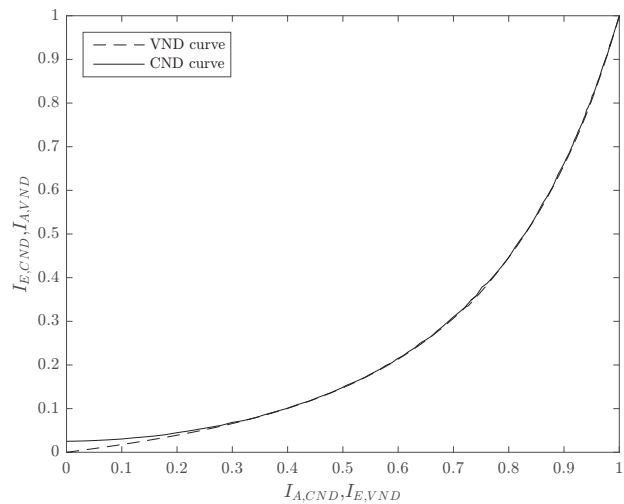}
\caption{EXIT Chart of optimized degree distributions for the rate $\frac{2}{3}$ multi-dimensional IRA lattice code.}
\label{fig:3}
\end{figure}
In our design, the portion of degree 1 CN must not be too small in order to ensure the decoder works in the first few iterations because our codes are nonsystematic \cite{Brink03}. From Fig. \ref{fig:3}, we can see that the VND curve literally touches the CND curve for the range $[0,1]$, which guarantees successful convergence and accurate decoding threshold.

We have adopted the proposed approach in designing the $\left(\alpha, \beta, 1+2i,\mathbb{H} \right)$-lattice ensemble with three code rates $\frac{3}{4}$, $\frac{2}{3}$ and $\frac{1}{2}$. The degree distributions and the decoding thresholds are shown in Table \ref{Table2}.

\begin{table*}
\caption{Optimal degree distributions and decoding thresholds of $\left(\alpha, \beta, 1+2i,\mathbb{H} \right)$-lattice ensemble with various code rates}
\label{Table2}
\centering
\begin{tabular}{|c|c|c|c|}
\hline
 Rates & Thresholds & Degree Distributions $(i,\alpha_i)$ for variable nodes, $(j,\beta_j)$ for check nodes \\ \hline
  $\frac{3}{4}$ & 4.47 dB & \begin{tabular}{c}
  $\alpha$: (2,0.288274), (3,0.265333), (7,0.188119), (13,0.123885), (15,0.134389)\\
  $\beta$: (1,0.055556), (3,0.944444)\\
  \end{tabular}  \\ \hline
 $\frac{2}{3}$ & 3.31 dB & \begin{tabular}{c}
  $\alpha$: (2,0.240605), (3,0.231215), (7,0.081754), (8,0.190942), (19,0.175951), (20,0.079534)\\
  $\beta$: (1,0.053861), (3,0.946139)\\
\end{tabular}  \\ \hline
 $\frac{1}{2}$ & 1.26 dB & \begin{tabular}{c}
  $\alpha$: (2,0.163689), (3,0.170788), (8,0.120858), (9,0.148837), (19,0.038618), (20,0.088323), (34,0.268886)\\
  $\beta$: (1,0.054328), (3,0.945672)\\
\end{tabular}    \\ \hline
\end{tabular}
\end{table*}
As shown in the table, the optimized CN distributions are degree 1 and degree 3 because this pair of CN distributions have the lowest optimization complexity and the minimum decoding threshold for the three code rates. We have also designed our codes with other pairs of CN distributions, but their performance is not much better than the code with only degree 1 and degree 3 CNs.

\begin{table*}[]
  \centering
 \caption{Comparisons of coding schemes}\label{table3}
\begin{tabular}{|c|c|c|c|}
\hline
 Coding schemes  & n [symbols]   & Coding loss [dB] & Gap to unrestricted Shannon limit [dB] \\  \hline
 GLD lattices \cite{Boutros14}  & 1,000  & 1.3  &  N/A\\  \hline
 \multirow{2}{*}{LDA lattices \cite{diPietro16}}  & 1,000  & 1.36 &  N/A \\
 \hhline{~---}
   & 10,000  & 0.7 &  N/A \\ \hline
  \multirow{3}{*}{LDA lattices  \cite{Boutros16}}  & 10,008  & 0.55 &  1.05 \\
  \hhline{~---}
    & 100,008  & 0.36 &  0.9 \\
    \hhline{~---}
   & 1,000,008  & 0.3 &  0.8 \\ \hline
 \multirow{3}{*}{ LDLCs \cite{Sommer06} }       & 1,000  & 1.5  & N/A \\
   \hhline{~---}
          & 10,000  & 0.8  & N/A \\
  \hhline{~---}
          & 100,000  & 0.6  & N/A \\ \hline
\multirow{2}{*}{ QC-LDPC lattices \cite{Khodaiemehr17}}  & 1,190  &  2 & N/A \\
  \hhline{~---}
    & 30,000  &  1.5 & N/A \\ \hline
\multirow{3}{*}{    IRA lattices } & 1,000  & 1.5 &  1.7 \\
  \hhline{~---}
      & 10,000  & 0.6 &  0.8 \\
  \hhline{~---}
      & 100,000  & 0.3 &  0.46 \\ \hline
\end{tabular}
\end{table*}

\section{Simulation Results}
\begin{figure}[ht!]
	\centering
\includegraphics[scale=0.46]{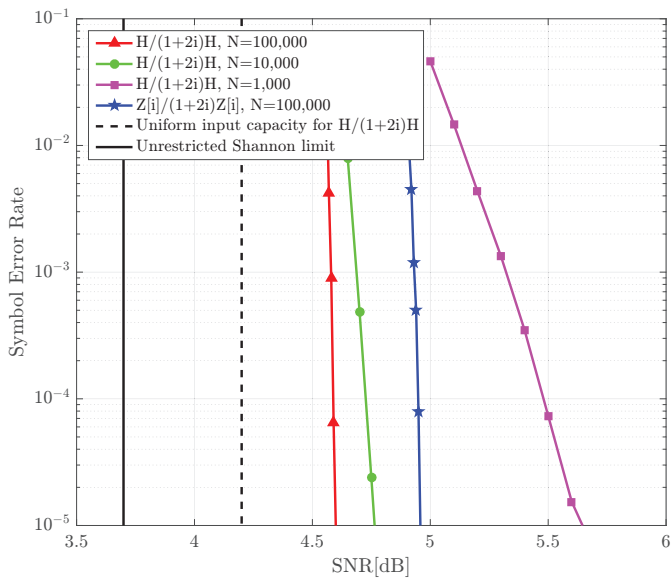}
\caption{Symbol error rate performance of rate $\frac{3}{4}$ codes.}
\label{fig:4}
\end{figure}

\begin{figure}[ht!]
	\centering
\includegraphics[scale=0.46]{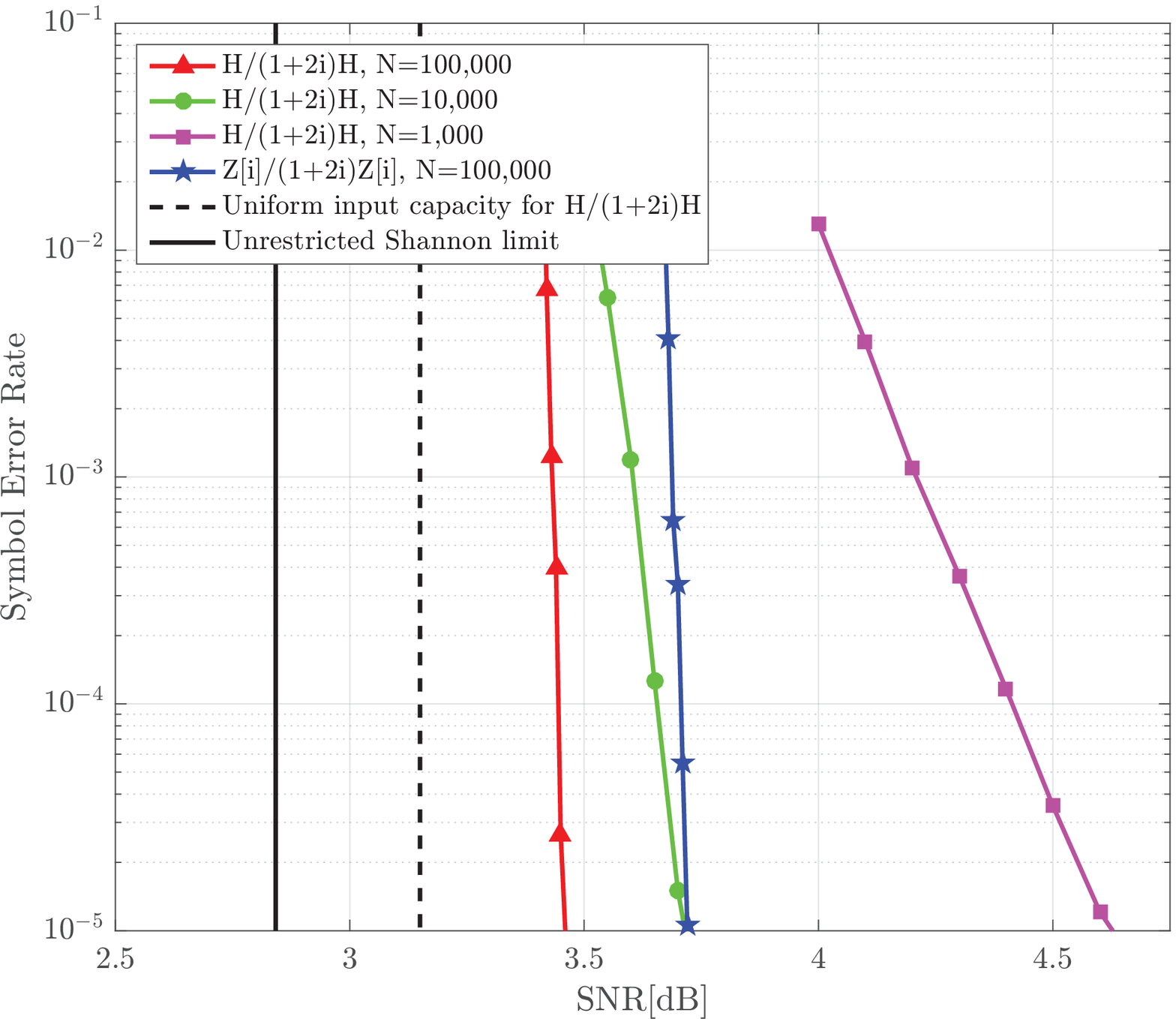}
\caption{Symbol error rate performance of rate $\frac{2}{3}$ codes.}
\label{fig:5}
\end{figure}

\begin{figure}[ht!]
	\centering
\includegraphics[scale=0.46]{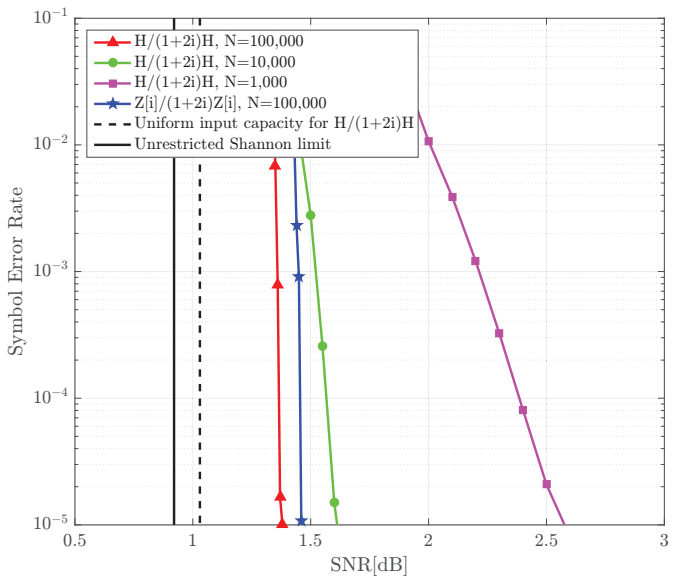}
\caption{Symbol error rate performance of rate $\frac{1}{2}$ codes.}
\label{fig:6}
\end{figure}

In this section, we present our simulation results for our multi-dimensional IRA lattice codes over $\mathbb{H}/(1+2i)\mathbb{H}$. In order to evaluate the average behavior of our codes, we randomly generated a codeword from the $(\alpha, \beta, 1+2i,\mathbb{H})$ ensemble and randomly select the values for $\mathbf{g}$, $\mathbf{g^\prime}$, $\mathbf{g^{\prime\prime}}$ and $\mathbf{r}$ in every channel realization. Since our coding scheme is based on \emph{finite constellations} with power constraint, the performance for three designed code rates $\frac{3}{4}$, $\frac{2}{3}$ and $\frac{1}{2}$ is measured in terms of symbol error rate (SER) versus SNR, which are depicted in Fig. \ref{fig:4}, Fig. \ref{fig:5} and Fig. \ref{fig:6}, respectively. Based on these designed code rates, the corresponding information rates are calculated by using (\ref{eq:aa3}) as $R_1 = 1.741$ bits/s/Hz, $R_2 =1.548$ bits/s/Hz and $R_3 =1.161$ bits/s/Hz, respectively. The corresponding unrestricted Shannon limit and uniform input capacity for each information rate are plotted in each figure. Additionally, we also show the SER performance for the previously designed IRA lattice codes over $\mathbb{Z}[i] /(1+2i) \mathbb{Z}[i]$ in all the figures for comparison because both partitions result in the same information rate. In our simulations, we set the codeword length to be 1,000, 10,000 and 100,000 symbols whereas the corresponding step sizes for SNR are 0.1 dB, 0.05 dB and 0.01 dB, respectively. The maximum number of decoding iterations was set to be 200.


In Fig. \ref{fig:4}, the unrestricted Shannon limit for $R_1$ is 3.70 dB. In this case, we observe that the gap to the unrestricted Shannon limit at the SER of $10^{-5}$ is 0.90 dB for our rate $\frac{3}{4}$ $D_4$-partition-based lattice code and 1.28 dB for the code in \cite{Qiu16}. Thus, our newly designed four-dimensional IRA lattice code is 0.38 dB better than the lattice code with two-dimensional lattice partitions. The unrestricted Shannon limit for $R_2$ is 2.84 dB. As shown in Fig. \ref{fig:5}, the gap between our lattice code and the unrestricted Shannon limit is 0.62 dB. For the code in \cite{Qiu16}, the gap is 0.88 dB. Therefore, the proposed lattice code is 0.26 dB better. Fig. \ref{fig:6} shows that the gap to the unrestricted Shannon limit is further reduced to 0.46 dB for our rate $\frac{1}{2}$ four-dimensional IRA lattice code. Our code is 0.1 dB better than the rate $\frac{1}{2}$ two-dimensional lattice code in \cite{Qiu16}. To this end, our proposed codes have lower decoding thresholds than that of the codes in \cite{Qiu16} but with higher encoding and decoding complexities.

Now we compare our designed lattice codes with the lattice coding schemes from \cite{Boutros16,Boutros14,diPietro16,Sommer06,Khodaiemehr17} for the same codeword length. Since these schemes are based on infinite constellations, their performances are measured in terms of gap to the Poltyrev limit which can be considered as coding loss \cite[Section VI-B]{Boutros16}. To obtain the coding loss in our lattice coding scheme, we measure the gap to uniform input capacity. The comparisons are listed in Table \ref{table3}, showing the simulation results which are reported for each scheme in the appropriate reference, including codeword length and coding loss when SER is at $10^{-5}$.

From Figs. \ref{fig:4}-\ref{fig:6}, one can observe that our code with rate $\frac{1}{2}$ have the smallest \emph{coding loss}. To be more specific, the coding loss for our lattice codes with $N=100,000$, $N=10,000$ and $N=1,000$ when SER is at $10^{-5}$ is about 0.3 dB, 0.6 dB and 1.5 dB. From Table \ref{table3}, it can be seen that our coding scheme outperforms all of these schemes for large codeword length, i.e., $N \geq 10,000$. When the codeword length is 1,000, our code is about 0.2 dB worse compared with LDA lattices \cite{diPietro16} and GLD lattices \cite{Boutros14} because of the probability of short cycles are higher when the codeword length is small. Since our goal is to design capacity-approaching lattice codes, thus we mainly focus on the codes with large codeword length, i.e., $N \geq 10,000$. Note that the direct comparison of encoding and decoding complexities for lattice codes with infinite constellations and our codes with finite constellations may not be fair and thus is omitted.

It is also worth noting that the waterfall regions of our multi-dimensional lattice codes are within 0.14 dB to the predicted decoding thresholds as shown in Table \ref{Table2} for various code rates. Therefore, it is evident that the proposed EXIT chart analysis for our multi-dimensional lattice codes is effective.

\section{Conclusion}
In this paper, we designed new multi-dimensional IRA lattice codes with finite constellations. Most compellingly, we proposed a novel encoding structure and proved that our codes can attain the permutation-invariance and symmetry properties in the densities of the decoder's messages. Under these properties, we used two-dimensional EXIT charts to analyze the convergence behavior of our codes and to minimize the decoding threshold. Our design can employ any higher-dimensional lattice partitions. Numerical results show that our designed and optimized lattice codes can achieve within 0.46 dB of the unrestricted Shannon limit and outperform existing lattice coding schemes for large codeword length.

\appendices
%

\newcounter{TmpEqCnt0} \setcounter{TmpEqCnt0}{\value{equation}} %
\setcounter{equation}{64}
\begin{figure*}[t]
\begin{align}\label{eq:lp1}
x_n[\tau] \oplus x_n[\upsilon] =& \left(\bigoplus\nolimits_{i=0}^{j_n-1}z_{a_n+i}[\tau]  \oplus c_{n-1}[\tau]\right) \oplus \left(\bigoplus\nolimits_{i=0}^{j_n-1}z_{a_n+i}[\upsilon]  \oplus c_{n-1}[\upsilon] \right) \oplus C_{gn} \oplus C_{gn} \nonumber \\
=&\left(\bigoplus\nolimits_{i=0}^{j_n-1} \left(z_{a_n+i}[\tau] \oplus z_{a_n+i}[\upsilon] \right) \oplus \left(c_{n-1}[\tau]\oplus c_{n-1}[\upsilon] \right) \right) \oplus C_{gn} \oplus C_{gn}
\end{align}
\hrule \vspace{-5mm}
\end{figure*}
\setcounter{equation}{\value{TmpEqCnt0}}

\section{Proof of Proposition \ref{linear}}
\label{appendix:1}
We divide our encoder into two parts: the first part is from the input of the repeater to the output of the interleaver; the second part is from the input of the combiner to the output of the accumulator. To prove that our codes are linear codes, we only need to show that the second part is a linear system. This is because the first part is already linear.

A linear code has the linear property such that the linear combination of two codewords is still a valid codeword. Now suppose we have two different codewords $\mathbf{X}^{\tau}$ and $\mathbf{X}^{\upsilon}$ with length $N$. The linear combination of these two codewords is
 \begin{align}\label{eq:lc1}
 \mathbf{X}^{\tau} \oplus \mathbf{X}^{\upsilon}=& [x_1[\tau],x_2[\tau],\cdots,x_N[\tau]] \oplus \nonumber \\
  &[x_1[\upsilon],x_2[\upsilon],\cdots,x_N[\upsilon]] \nonumber \\
 =&[x_1[\tau]\oplus x_1[\upsilon] ,x_2[\tau]\oplus \nonumber \\
 &x_2[\upsilon],\cdots,x_N[\tau]\oplus x_N[\upsilon]],
 \end{align}
 where $\oplus$ is the modulo lattice addition. Now, we focus on the $n$-th component of the codeword $x_n$ for $ 1 \leq n \leq N$.
 The encoding function for the $n$-th component of the codeword is
 \begin{equation}
\left(\bigoplus\nolimits_{i=0}^{j_n-1}z_{a_n+i} \oplus g_{a_n+i} \right) \oplus c_{n-1} \oplus g_n^\prime \oplus g_n^{\prime\prime}  = x_n,
\end{equation}
where $z_{a_n}$ and $z_{a_n+j_n-1}$ represent the first and last interleaved symbols to the $n$-th combiner; $c_{n-1}$ is the $n-1$-th output of the time-varying accumulator. Note that the random-coset is removed before iterative decoding, thus it is not considered as part of the codebook information.

We can then rewrite the above equation as
\begin{equation}
\bigoplus\nolimits_{i=0}^{j_n-1}z_{a_n+i}  \oplus c_{n-1} \oplus C_{gn} = x_n,
\end{equation}
where $\bigoplus_{i=0}^{j_n-1}g_{a_n+i} \oplus g_n^\prime \oplus g_n^{\prime\prime} = C_{gn} \in \Psi$ and $C_{gn}$ is the constant associated with $x_n$. Note that the term $\bigoplus_{i=0}^{j_n-1}g_{a_n+i}$ can be extracted by using the associative law on the addition of Hurwitz integers.

Now for the $n$-th codeword component in $\mathbf{X}^{\tau}$ and $\mathbf{X}^{\upsilon}$, we have
\begin{equation}
\bigoplus\nolimits_{i=0}^{j_n-1}z_{a_n+i}[\tau]  \oplus c_{n-1}[\tau] \oplus C_{gn} = x_n[\tau].
\end{equation}
\begin{equation}
\bigoplus\nolimits_{i=0}^{j_n-1}z_{a_n+i}[\upsilon]  \oplus c_{n-1}[\upsilon] \oplus C_{gn} = x_n[\upsilon].
\end{equation}
Here $C_{gn}$ is \emph{deterministic} for a particular codeword position. The linear combination in Eq. (\ref{eq:lc1}) becomes Eq. (\ref{eq:lp1}) for $ 1 \leq n \leq N$, which is shown at the top of the page. The deterministic part $C_{gn} \oplus C_{gn}$ can contribute to non-linearity when $C_{gn} \oplus C_{gn} \neq C_{gn}$.
Therefore, when we let $C_{gn} = 0$, our codes are linear.

\section{Proof of Theorem \ref{OS}}\label{appendix:4}
Consider the $n$-th symbol. Let $X_n$ be the channel input random variable. Let $Y_n$ be the $n$-th received signal with the input-output relationship given by
\setcounter{equation}{65}
\begin{equation}
Y_n = X_n+N_n  \overset{(b)}= C_n \oplus R_n+N_n,
\end{equation}
where $N_n\sim \mathcal{N}(0,\sigma_{ch}^2)$ is the noise of the AWGN channel; (b) follows Eq. (\ref{eq:40}); $C_n$ is the $n$-th random variable of intended codeword before adding the random-coset and $R_n$ is the $n$-th random variable of the random-coset.

To prove that adding the random-coset can produce the output-symmetric effect, we must have
\begin{equation}\label{eq:ss1}
\text{Pr}[Y_n \not\in \mathcal{U}(X_n)| C_n = \psi_i] = \text{Pr}[Y_n \not\in \mathcal{U}(X_n)| C_n = \psi_j],
\end{equation}
where $\mathcal{U}(.)$ outputs the maximum-likelihood decision region; $\psi_i,\psi_j \in \Psi$ and $\psi_i \neq \psi_j$. In other words, the decoding error probability is the same for any transmitted codeword.

For the left term in Eq. (\ref{eq:ss1}), we have
\begin{align}
\text{Pr}&[Y_n \not\in \mathcal{U}(X_n)| C_n=  \psi_i] \nonumber \\
&= \sum_{r_i} \text{Pr}[Y_n \not\in \mathcal{U}(X_n) | R_n = r_i, C_n=\psi_i]\times \nonumber \\
& \;\;\;\;\;\;\;\;\;\;\; \text{Pr}[R_n = r_i|C_n=\psi_i].
\end{align}
Since $R_n$ is independent of $C_n$ and $R_n$ is uniformly distributed over $\Psi$, we then have
\begin{align}\label{eq:new1}
&\text{Pr}[Y_n \not\in \mathcal{U}(X_n)| C_n = \psi_i]  \nonumber \\
&= \sum_{r_i} \text{Pr}[Y_n\not\in \mathcal{U}(X_n) | R_n = r_i, C_n=\psi_i]\cdot \text{Pr}[R_n = r_i] \nonumber \\
& = \sum_{x_i} \text{Pr}[Y_n \not\in \mathcal{U}(X_n)| X_n = x_i = r_i \oplus \psi_i ]\cdot \text{Pr}[R_n = r_i] \nonumber \\
& = \frac{1}{p^M} \sum_{x_i} \text{Pr}(Y_n \not\in \mathcal{U}(X_n)| X_n = x_i).
\end{align}
Similarly, for a different realisation of $C_n$ and $R_n$, we have
\begin{align}\label{eq:new2}
&\text{Pr}[Y_n \not\in \mathcal{U}(X_n)| C_n = \psi_j]  \nonumber \\
& =\sum_{r_j} \text{Pr}[Y_n\not\in \mathcal{U}(X_n) | R_n = r_j, C_n=\psi_j]\cdot \text{Pr}[R_n = r_j] \nonumber \\
& = \sum_{x_j} \text{Pr}[Y_n \not\in \mathcal{U}(X_n)| X_n = x_j = r_j \oplus \psi_j ]\cdot \text{Pr}[R_n = r_j] \nonumber \\
& = \frac{1}{p^M} \sum_{x_j} \text{Pr}(Y_n \not\in \mathcal{U}(X_n)| X_n = x_j).
\end{align}
Since the ranges of $x_i$ and $x_j$ are $\Psi$, therefore we can obtain that
\begin{align}\label{eq:ssl1}
\sum_{x_i} \text{Pr}(Y_n \not\in \mathcal{U}(X_n)& | X_n = x_i) \nonumber \\
=& \sum_{x_j} \text{Pr}(Y_n \not\in \mathcal{U}(X_n)| X_n = x_j).
\end{align}
Plugging Eq. (\ref{eq:ssl1}) into Eq. (\ref{eq:new1}) and Eq. (\ref{eq:new2}), we obtain Eq. (\ref{eq:ss1}).

\section{Proof of Permutation-Invariance}
\subsection{Proof of Theorem \ref{PERI}}\label{appendix:2}
First, we define a probability-vector random variable $\mathbf{X} = [X_{\psi_0},X_{\psi_1},\ldots,X_{\psi_{p^M-1}}]$ and let $\mathbf{P} = \mathbf{X}^{+\theta}$ where $\theta$ is a random variable and uniformly chosen from $\Psi$. For the $m$-th random variable in $\mathbf{X}$, we denote a probability event by
\begin{equation}
\text{Pr}[X_{\psi_m} \in \varepsilon].
\end{equation}

Then for the $i$-th random variable in $\mathbf{P}$, we have
\begin{equation}
\text{Pr}[P_{\psi_i} \in \varepsilon] = \text{Pr}[X_{\psi_m} \in \varepsilon] \cdot \text{Pr}[\psi_m \oplus \theta = \psi_i],
\end{equation}
because $\theta$ is independent of $\mathbf{X}$.

Similarly, for the $j$-th random variable in $\mathbf{P}$, where $\psi_j \neq \psi_i$, we can obtain that:
\begin{equation}
\text{Pr}[P_{\psi_j} \in \varepsilon] = \text{Pr}[X_{\psi_m} \in \varepsilon] \cdot \text{Pr}[\psi_m \oplus \theta = \psi_j],
\end{equation}

We know $\theta$ is a random variable and uniformly chosen from $\Psi$. Thus we have:
\begin{equation}
\text{Pr}[\psi_m \oplus \theta = \psi_i] = \text{Pr}[\psi_m \oplus \theta = \psi_j] = \frac{1}{p^M}.
\end{equation}

Therefore, the distribution of any two random variables in $\mathbf{P}$ is the same. If we let $\psi_j = \psi_i \oplus \chi$ for any fixed $\chi \in \Psi$, we obtain that:
\begin{equation}
\text{Pr}[P_{\psi_i} \in \varepsilon] = \text{Pr}[P_{\psi_j} \in \varepsilon] = \text{Pr}[P_{\psi_i \oplus \chi} \in \varepsilon]= \text{Pr}[P_{\psi_i}^{ \oplus \chi} \in \varepsilon].
\end{equation}
It can be seen that every random variable in $\mathbf{P}$ is identically distributed. Therefore, we can conclude that $\mathbf{P}$ is identically distributed with $\mathbf{P}^{ \oplus \chi}$ so $\mathbf{P}$ is permutation-invariant.

\subsection{Proof of Lemma \ref{PERIW}}\label{appendix:3}
For the $m$-th LLR random variable in $\mathbf{W}$, we denote a probability event by
\begin{equation}
\text{Pr}[W_{\psi_m} \in \delta],
\end{equation}
where $\delta$ is a random event.
From (\ref{eq:LLR1}), we know that $W_{\psi_m} = \log \left( \frac{P_{\psi_0}}{P_{\psi_m}}\right)$, thus we can obtain that
\begin{align}\label{eq:pi1}
\text{Pr}&[W_{\psi_m} \in \delta] \nonumber \\
&= \text{Pr}\Bigg[\log \left( \frac{P_{\psi_0}}{P_{\psi_m}}\right) \in \delta \Bigg] \nonumber \\
& = \text{Pr}[ P_{\psi_0} \in e^\delta P_{\psi_m} ] \nonumber \\
& = \int_{P_{\psi_m}}\int_{e^\delta P_{\psi_m}} f_{P_{\psi_0},P_{\psi_m}}(p_{\psi_0},p_{\psi_m}) dp_{\psi_0} \, dp_{\psi_m},
\end{align}
where $f_{P_{\psi_0},P_{\psi_m}}(p_{\psi_0},p_{\psi_m})$ denotes the joint pdf of $P_{\psi_0}$ and $P_{\psi_m}$.

Similarly, for the $n$-th LLR random variable in $\mathbf{W}$ where $n \neq m$, we have
\begin{align}\label{eq:pi2}
\text{Pr}[W_{\psi_n} \in \delta]  = \int_{P_{\psi_n}}\int_{e^\delta P_{\psi_n}} f_{P_{\psi_0},P_{\psi_n}}(p_{\psi_0},p_{\psi_n}) dp_{\psi_0} \, dp_{\psi_n}.
\end{align}

We know $P_{\psi_m}$ and $P_{\psi_n}$ have the same distribution as because $\mathbf{P}$ is permutation-invariant. Thus, the joint distribution of $P_{\psi_0}$ and $P_{\psi_m}$ is the same as that of $P_{\psi_0}$ and $P_{\psi_n}$. As a result, we can obtain that:
\begin{equation}
\text{Pr}[W_{\psi_n} \in \delta] = \text{Pr}[W_{\psi_m} \in \delta].
\end{equation}
This indicates that $W_{\psi_n}$ and $W_{\psi_m}$ have the same distribution for any $n \neq m$. Therefore, $\mathbf{W}$ is permutation-invariant.

\bibliographystyle{IEEEtran}
\bibliography{journal_abbr,IRA_HD}

\end{document}